\documentclass[aps,prl,groupedaddress,twocolumn]{revtex4}
\usepackage{graphicx}
\begin{document}
\draft

\title{Spatial Determination of Magnetic Avalanche Ignition Points}
\author{Reem  Jaafar, S. McHugh, Yoko Suzuki, and M. P. Sarachik}
\address{Physics Department, City College of the City University of New York, New York, NY 10031}
\author{Y. Myasoedov, E. Zeldov, and H. Shtrikman}
\address{Dept.\,of Condensed Matter Physics, The Weizmann Institute of Science, Rehovot
76100, Israel}
\author{R. Bagai and G. Christou}
\address{Department of Chemistry, University of Florida, Gainesville, FL 32611}
\date{\today}

\begin{abstract}
Using time-resolved measurements of local magnetization in the molecular magnet Mn12-ac, we report studies of the propagation of magnetic avalanches (fast magnetization reversals) that originate from points inside the crystals rather than at the edges.  The curved nature of the fronts produced by avalanches is reflected in the time-of-arrival at micro-Hall sensors placed at the surface of the sample.  Assuming that the avalanche interface is a spherical bubble that grows with a radius proportional to time, we are able to locate the approximate ignition point of each avalanche in a two-dimensional cross-section of the crystal.  For the samples used in these studies, avalanches in a given crystal are found to originate in a small region with a radius of roughly $150 \mu m$.

\end{abstract}

\pacs{PACS numbers: 71.30.+h, 73.40.Qv, 73.50.Jt}

\maketitle

Mn$_{12}$-acetate, or [Mn$_{12}$O$_{12}$(CH$_3$COO)$_{16}$(H$_2$O)$_4$]$\cdot$2CH$_3\\$COOH$\cdot$4H$_2$O] (hereafter referred to as Mn$_{12}$-ac), is a prototypical molecular magnet composed of magnetic molecules with cores consisting of twelve Mn atoms strongly coupled by exchange to form superparamagnetic clusters of large spin $S = 10$ at low temperatures \cite{Lis},\cite{Sessoli}.  Arranged in a centered tetragonal lattice, the Mn$_{12}$ clusters are subject to strong magnetic anisotropy along the symmetry axis (the c-axis of the crystal).  Below the blocking temperature of about 3.5 K, the crystal exhibits remarkable staircase magnetic hysteresis due to resonant quantum spin tunneling between energy levels on opposite sides of the anisotropy barrier corresponding to different spin projections along the easy axis\cite{Friedman}.

Recent local time-resolved measurements of fast magnetization reversal in single crystals of Mn$_{12}$-ac have indicated that a magnetization avalanche spreads as a narrow interface that propagates through the crystal at a constant velocity that is roughly two orders of magnitude smaller than the speed of sound \cite{Suzuki}.  This has been attributed to "magnetic deflagration", in analogy with the propagation of a flame front through a flammable chemical substance, a phenomenon called chemical deflagration \cite{deflagration}.  In the earlier study \cite{Suzuki}, the avalanches occurred stochastically in response to a time-varying (swept) magnetic field applied along the easy axis of the Mn$_{12}$-ac crystals.  Under the conditions and for the samples used, avalanches invariably originated at or near the long end of the crystal, and the magnetic front detected by sensors placed near the middle detected the propagation of a planar (or approximately planar) propagating front.

In this paper we report a study of avalanches triggered in a swept magnetic field at points within the interior of the crystals rather than at the edges.  Assuming that the velocity of propagation of the avalanche is isotropic and gives rise to a spherical propagating front, the time-of-arrival at sensors placed at the surface of the sample allowed us to trace each avalanche back to its approximate point of ignition. For the crystals used in the current experiment, the avalanches typically originate in some small region within a crystal.

The measurements are described in detail in reference \cite {Suzuki}. Small Hall bars (six or seven, depending on the sample) of dimensions $30 \times 30$ $\mu$m$^2$ placed at $80$ $\mu$m (center to center) intervals were used to measure the magnetization of single crystals of Mn$_{12}$-ac in the form of parallelepipeds of typical cross section $0.3 \times 0.3$ mm$^2$ and length $1$ to $1.4$ mm.  Operating at $0.255$ K, avalanches were recorded for magnetic fields between $1.5$ and $3.5$ T.  The Hall sensors were aligned to detect the magnetic induction of the sample in the $x$-direction, $B_x$, which is proportional to the spatial derivative of $M_z$ in the  region near the sensor, $B_x \propto \partial M_z/\partial z$.\cite {avraham}  $B_x$ is proportional $M_z$, and for a uniform magnetization in the z-direction,  derives mostly from the strong gradient at the  sample ends.  During an avalanche, there is a large contribution to $B_x$ at the front, the narrow propagating interface between regions of opposing magnetization, where $\partial M_z/\partial z$ is large.  Each sensor thus registers a large signal as the front passes by.  The sign of the peak, positive or negative, is determined by the direction of the field lines.

Figure 1 shows the time evolution of an avalanche starting from some region away from the Hall sensors near the end of the sample.  The schematic (top center) shows the magnetization of the crystal and the field lines that give rise to the observed signal.  The graphed inset shows the peak position of the signal plotted as a function of time; the slope of the straight line yields the velocity of propagation of the front.  These data are consistent with the results of reference\cite {Suzuki}.


\begin{figure}
\includegraphics[width=1.0\columnwidth]{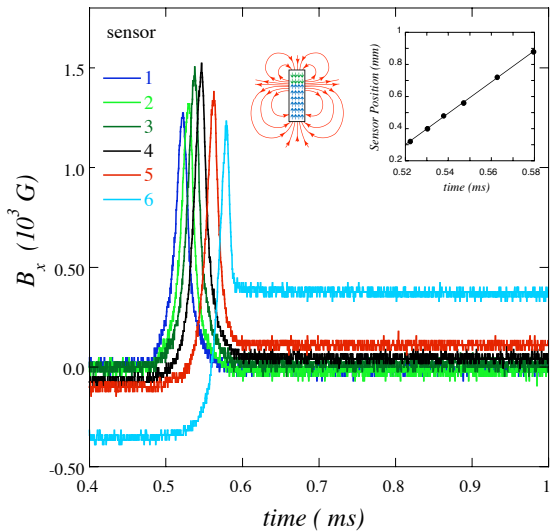}
\caption{Signals recorded during a magnetic avalanche by Hall sensors placed on the surface near the middle of a bar-shaped sample of Mn$_{12}$-acetate.  The schematic indicates the field line configuration that gives rise to the signals as the avalanche propagates from one end of the sample to the other.  The inset shows sensor positions versus the time at which each sensor recorded the peak amplitude for an avalanche that started at the top and travels downward.
\label{fig1}}
\end{figure}

Figure 2 shows the sensor traces for an avalanche observed in a different sample that produces both positive and negative signals.   As before, we plot the positions of the peaks as a function of time for two such avalanches in Fig. 3.  In contrast with the straight line shown in the inset to Fig. 1, which we have attributed to the planar front of an avalanche propagating from the end of the sample,  we see a "V-shaped" curve with negative and positive curved branches.  This can be understood as an avalanche starting near the middle of the sample and propagating in opposite directions towards the sample ends.  As illustrated in the schematic inset in Fig. 2, the signals of opposite sign are due to field lines pointing in opposite directions corresponding to the upper (lower) front traveling upward (downward).  Figure 3(a) shows an avalanche initiating above the third Hall sensor while Fig. 3(b) shows one originating between the fourth and fifth.  The avalanches of Fig. 3 clearly show curvature absent from the planar front of Fig. 1 and are distinctly different from one another.  We propose that the curvature is determined by the distance from the ignition point to the surface on which the Hall sensors are mounted.   This can be understood by simple geometric considerations.


\begin{figure}
\includegraphics[width=1.0\columnwidth]{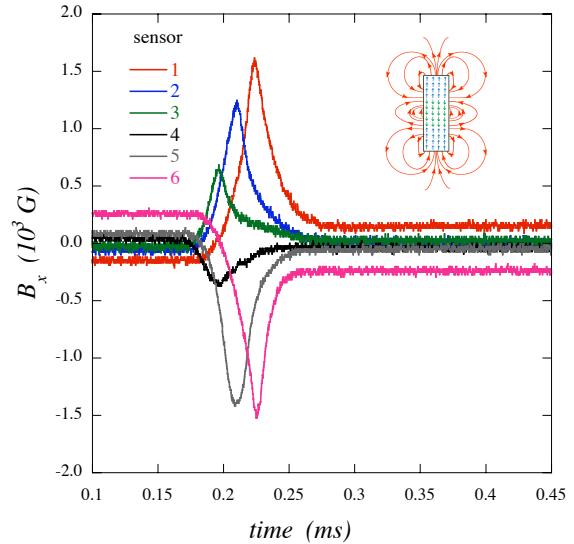}
\caption{Sensor signals for an avalanche starting near the middle of the sample.  The schematic indicates the (quadrupolar) field configuration associated with avalanche fronts propagating in opposite directions toward the two ends, giving rise to the observed positive-going and negative-going signals. 
\label{fig2}}
\end{figure}

We assume the avalanche grows as a spherical bubble with a radius growing linearly in time, $r=vt$.  In other words,
\[v^2t^2 = (x-x_0)^2 + (y-y_0)^2 + (z - z_0)^2,\]
where  $(x_0, y_0, z_0)$ is the ignition point within the sample.  
With Hall bars along the z-axis and $y=x=0$, the position of the peak as a function of time is given as the intersection of the spherical bubble with the z-axis,
\begin{eqnarray}
z(t) = z_0 \pm \sqrt{v^2t^2 + d^2},
\end{eqnarray}
where $d$ is the distance $\sqrt{(y_0)^2 + (x_0)^2}$ between the point of ignition and the z-axis containing the Hall sensor array.  


\begin{figure}
\includegraphics[width=1.0\columnwidth]{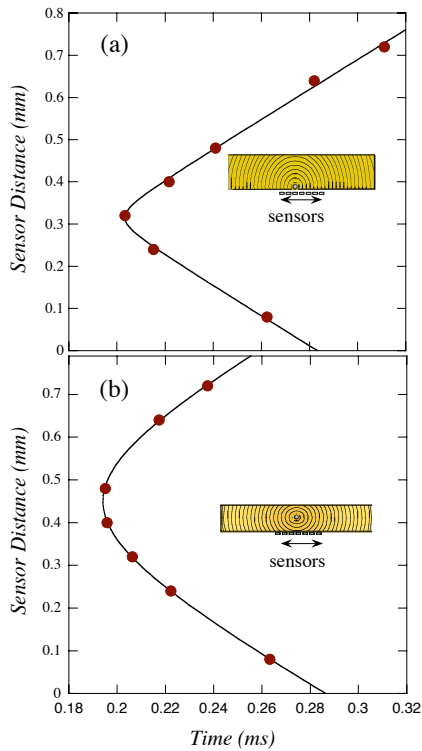}
\caption{For two avalanches, the sensor position is shown as a function of the time of arrival of the peak amplitude at each sensor.  In contrast with the inset shown in Fig. 1, the traces are not straight  lines, and the curvature is different for the avalanches shown in frame (a) and frame (b).  As shown in the text, frame (a) corresponds to an avalanche triggered near the surface while frame (b) shows an avalanche triggered far from the surface on which the sensor was mounted.
\label{fig3}}
\end{figure}

By fitting the peak positions in time with this expression, we can extract values of $z_0$ and $d$, and thereby determine the point where the avalanche started.  Thus, for example, frames (a) and (b) of Fig. 3 show data with distinctly different curvatures for two different avalanches originating within the bulk of the same sample.  As illustrated in the schematic diagrams, the smaller curvature of Fig. 3a is caused by an avalanche triggered near the surface on which the sensors are mounted, while the larger curvature of Fig. 3b is due to an avalanche originating further away.

For a particular sample of dimensions $1.4 \times 0.3 \times 0.3 $ mm$^3$, Fig. 4 shows a distribution of the avalanche ignition points projected onto the $d-z$ plane.  Hall sensors were placed near the middle of the sample extending from $z  \approx 0.4$ to $z  \approx 1.0$ mm.  The vertical lines near the ends of the sample indicate "blind" regions outside the Hall sensor array where avalanches could not be traced back to their origin.  Only  $\approx 5$ \% of the avalanches were triggered in these regions.  For this sample, the ignition points lie predominantly within a region near the center of radius roughly $150\mu m$.

In most of the crystals we have studied, the avalanches originate at the ends.  This suggests that dipolar demagnetizing fields, or other magnetic effects associated with the surface, play a role in the ignition process.  In this paper we have reported the observation of avalanches that originate in a region within the bulk of a crystal.  We suggest that these bulk-initiated avalanches originate from "weak" regions of the crystal where the density of defects (or perhaps the second species known to exist in Mn12 ac) is higher.  The location and effect of these regions should vary from crystal to crystal and should depend on growing conditions.


\begin{figure}
\vspace{0.3in}
\includegraphics[width=1.0\columnwidth]{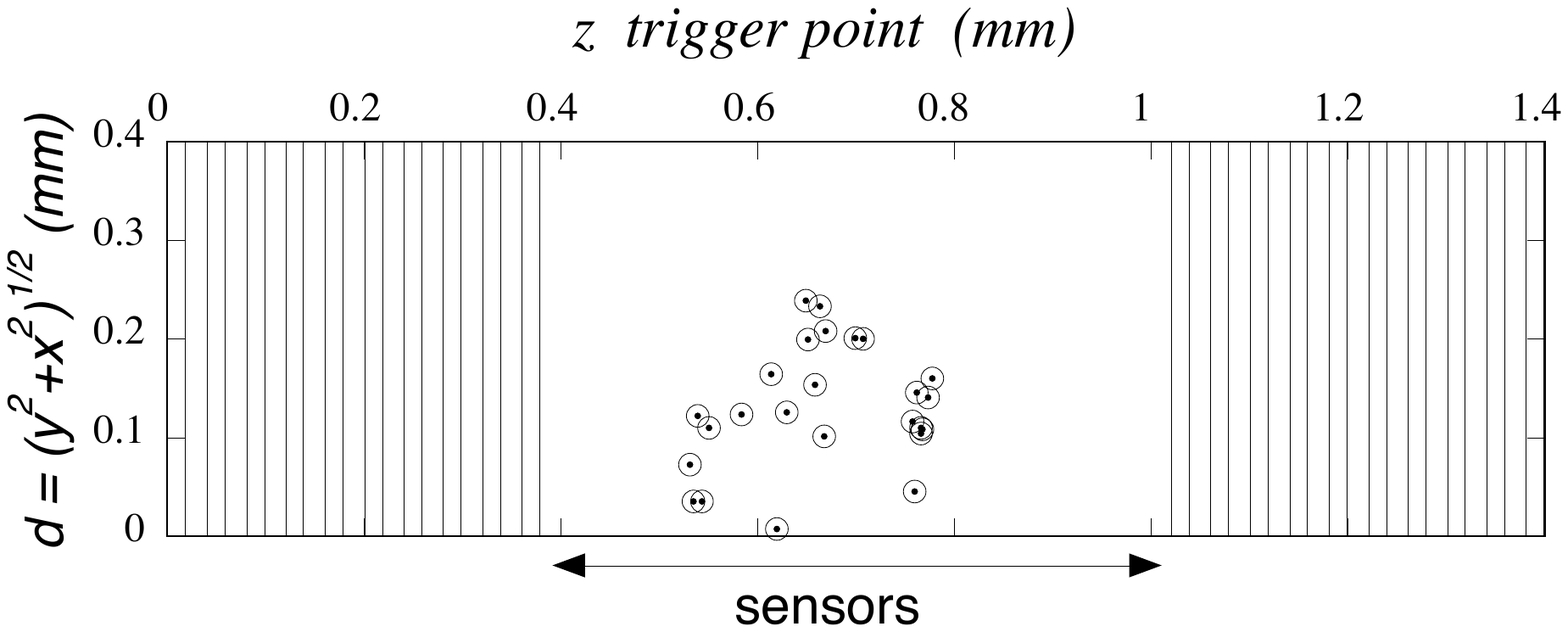}
\caption{Distribution of ignition points of avalanches for a particular sample projected onto the $d-z$ plane.  The double arrow indicates the region spanned by Hall sensors; the vertical lines designate the regions beyond the Hall sensor array where avalanches could not be traced to their points of origin.
\label{fig4}}
\end{figure}

This work was supported at City College by NSF grant DMR-00451605.  E. Z. and H. S. acknowledge
the support of the Israel Science Foundation Center of Excellence grant 8003/02.  Support for G. C. was provided by NSF grant CHE-0414555.

 \end{document}